\documentclass[letterpaper]{article}
\usepackage{aaai}
\usepackage{times}
\usepackage{helvet}
\usepackage{courier}
\usepackage{algorithm}
\usepackage{algpseudocode}
\usepackage{graphicx}

\usepackage[tight,footnotesize]{subfigure}
\graphicspath{{./Pictures/}}
\usepackage{etoolbox}
\AtBeginEnvironment{algorithmic}{\scriptsize}
\begin{document}

\title{Decentralised Multi-Agent Reinforcement Learning for Dynamic and Uncertain Environments}
\author{{Andrei Marinescu},
{Ivana Dusparic},
{Adam Taylor},
{Vinny Cahill},
{Siobh\'{a}n Clarke}
\\{Distributed Systems Group, School of Computer Science and Statistics}\\
{Trinity College Dublin, Ireland}\\
\textup{{\{marinesa, ivana.dusparic, tayloral, vinny.cahill, siobhan.clarke\}@scss.tcd.ie}
}
}
\maketitle
\begin{abstract}

Multi-Agent Reinforcement Learning (MARL) is a widely used technique for optimization in decentralised control problems. However, most applications of MARL are in static environments, and are not suitable when agent behaviour and environment conditions are dynamic and uncertain. Addressing uncertainty in such environments remains a challenging problem for MARL-based systems. The dynamic nature of the environment causes previous knowledge of how agents interact to become outdated. Advanced knowledge of potential changes through prediction significantly supports agents converging to near-optimal  control solutions. In this paper we propose P-MARL, a decentralised MARL algorithm enhanced by a prediction mechanism that provides accurate information regarding up-coming changes in the environment. This prediction is achieved by employing an Artificial Neural Network combined with a Self-Organising Map that detects and matches changes in the environment. The proposed algorithm is validated in a realistic smart-grid scenario, and provides a 92\% Pareto efficient solution to an electric vehicle charging problem.
\end{abstract}

\section{Introduction}
Multi-Agent Reinforcement Learning (MARL) is being increasingly used in various domains such as robotics, computer networks, traffic, resource management, robotic teams and distributed control in general \cite{Busoniu2008}. Many of these situations pose complex challenges to multi-agent systems due to the dynamicity of the environment, defined by uncertainty and non-stationary behaviour. Adding to the complexity in such circumstances is the situation where one might not only encounter dynamicity generated by the stochastic interactions between agents, but also stochasticity due to a possibly continuously changing and evolving environment (independently of agent actions). The environment's behavioural patterns are not being repeated, as the environment is continuously evolving. Moreover, sudden unexpected events which weren't encountered previously are contributing to the element of uncertainty. 

Reinforcement Learning (RL) \cite{Sutton1998} can provide an optimal solution for static environments, where a single agent perceives the current state of the environment and takes a decision which affects the environment. There is a finite number of changes in the environment in such cases. Despite these, RL does not require any previous knowledge of the environment. Through a reward based system, the agent eventually learns its optimal behaviour by trial-and-error, where the reward system provides feedback for each of the chosen actions (either a reward or a penalty depending on the case). The agent attempts to maximize its overall reward, and as a results it converges in the end to an optimal solution. The problem arises when several such RL based agents interact within the same environment, transforming (even initially) static environments (from a single agent's perspective) in non-stationary environments, as each agent has a particular influence over the environment itself \cite{Chalkiadakis2003}. All agents attempt to learn simultaneously in such situations, and as a result the guarantees of algorithms convergence are lost. Therefore we are faced with a much more complex problem.

We propose to tackle the problem of non-stationary environments by minimizing the uncertainty component of the problem in order to provide a solution to a \emph{close enough}  problem. We augment an online MARL algorithm with a prediction and pattern change detection module in order to mitigate uncertainty. While an initial environment estimate is provided by the prediction side of the model, it also closely monitors the real-time behaviour to check whether there are any significant changes from the expected one (these might occur due to increased uncertainty). If such changes happen, the module triggers the reprediction of the environment's future behaviour, based on previously encountered similar patterns. 

Our algorithm, Predictive-Marl (P-MARL), is tested in a real world smart-grid scenario, where a set of Electric Vehicles (EVs) are designated to optimally reach a desired battery state of charge before they are to depart from home. The optimality criteria is defined by the price of electricity, which in this case is considered directly proportional with the aggregated power demand. This is a decentralised control problem, where agents' action affect the state of the environment, and where even the underlying environment's initial state is not known ahead of time with certainty. While there are methods for predicting future energy demands, particular events can lead to unexpected demand patterns which increase the level of uncertainty of such a system. Such events can be caused by natural disasters, power network failures, and unexpected weather phenomena. These are not generally foreseeable.

In the following sections of this paper we introduce the background and related work with regard to the problem, we continue with the generic P-MARL proposition, while the next section presents an implementation on a specific example - a real world Smart Grid scenario. In the final sections we evaluate the algorithm's performance and present our conclusions with regard to its results.

\section{Background and Related Work}

 The problem faced is essentially a more complex version of the Distributed Constraint Optimization Problem (DCOP) \cite{Nguyen2014,Wu2013}. While optimal solutions exist for DCOP, these are NP-complete and are not suitable for large scale problems. More than that, a DCOP involving uncertainty \emph{does not have an optimal solution}, precisely due to the uncertainty involved in the environment's next state, which does not pertain a fully defined problem. This particular type of uncertain DCOP has also been defined as a Distributed Coordination of Exploration and Exploitation (DCEE) problem by \cite{Taylor2010}, as DCOP under Stochastic Uncertainty (StochDCOP) by \cite{Leaute2011} or simply dubbed DCOP with uncertainty \cite{Stranders2011}. Since the environment is uncertain and non-stationary, agent rewards will continuously change, therefore a trade-off  between exploration and exploitation is necessary in order to lead to a sufficiently good solution. As the environment's state is not known ahead of time, and is continuously undergoing change, the \emph{a priori} design of agent behaviour becomes infeasible.

Centralised solutions have been proposed to such problems, but there are issues concerning computational complexity (they are generally NP-complete), scalability of the solution, the privacy and independence of users, communication overhead involved and resilience/reliability of a centralised controller. Therefore decentralised options are more desirable for this kind of problem, and even more the ones that can adapt in real-time to new conditions produced by the uncertainty element of the environment. 

In the following section we present a novel algorithm, P-MARL - a decentralised reinforcement learning based approach to DCOP with uncertainty.

\section{System Architecture}
\label{sec:syst_arch}

Our chosen application environment is inherently non-stationary. In mathematical terms this means that the the underlying generating function \emph{f} of the environment changes over time. The environment effectively experiences \emph{concept drift} \cite{Schlimmer1986}- the environment in this case is also known as being dynamic. To simplify the problem one can attempt to model the outcome of the generating function of the environment, \emph{f}, based on recently observed behaviour. At every time step \emph{t} we have historical data available: $X^{H}=(X_{1},...,X_{t})$. We attempt to predict $X_{t+1}$ considering past observations. The core assumption when dealing with a concept drift problem is uncertainty about the future. Here our target $X_{t+1}$ is not known. It can be assumed, estimated or predicted, but there are no certainty guarantees. 

 To alleviate the concept drift problem we propose a two step solution, depending on the degree of \emph{driftness}, or the speed of drift:
\begin{enumerate}
\item The passive solution: a continuously updating model based on the most recently observed samples, as suggested by \cite{Widmer1996}
\item The active solution: once a pattern change detection mechanism notices an unexpected shift from the model's expected behaviour it triggers the generation of a new model - such as the solution proposed by \cite{Alippi2008}. 
\end{enumerate}

The second part of the solution is obviously more delicate. The new model is proposed based on a pattern matching mechanism that should be able to provide a close match, based on similarly previously encountered behaviour. This would in turn provide help in generating a better model. 

\begin{figure}[h]
\centering
\includegraphics[width=0.8\linewidth]{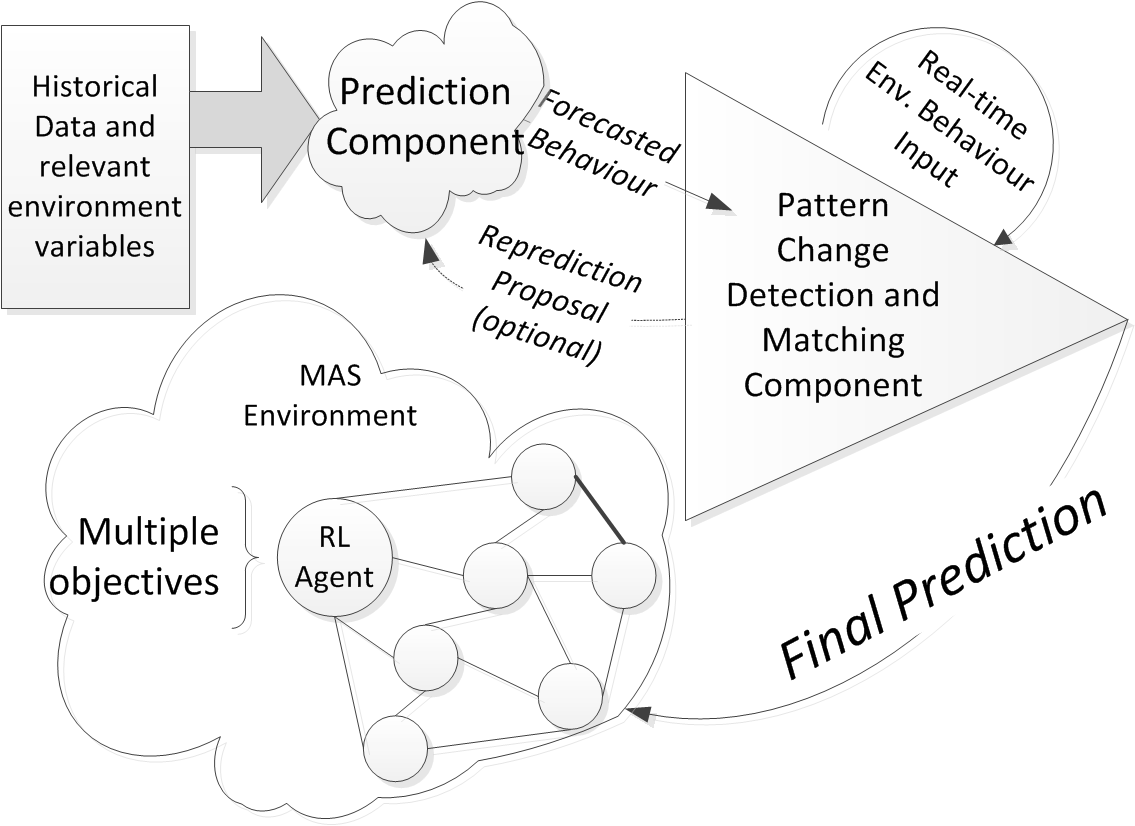} 
\caption[c]{MARL Algorithm Architecture}
\label{fig:alg_architecture}
\end{figure}

As a result, P-MARL's architecture comprises three key components:
\begin{enumerate}
\item \textbf{The Prediction Model}, which effectively considers recent historical values and other key variables that can affect the environment in order to provide an estimate of future behaviour. 
\item \textbf{The Pattern Change Detection and Matching Component}, which detects when the prediction model fails in providing reasonable estimations of the future state of the environment. This triggers a new model based subject to the latest observations and findings from a database. 
\item \textbf{The Multi-Agent System} (MAS) part based on \textbf{reinforcement learning}, that employs the previous components as an input to improve its performance in the dynamic environment. The RL side is implemented as a multi-objective W-learning process \cite{Humphrys1995}.
\end{enumerate}

The relationship between the components is illustrated in Fig. \ref{fig:alg_architecture}. Eventually, from the first two components, an estimate of the environment's future expected behaviour is provided. The MAS agents evaluate the future behaviour and attempt to optimally reach their goals with respect to the imposed constraints (these constraints can be implemented as extra objectives). 
In the following section we define the evaluation testbed environment, and we show in detail how P-MARL is implemented for such case.

\section{Smart Grid Case Study}

In the smart-grid environment, the model provides a good estimate of the next day's demand to the agents. The MAS agents in this case are EVs, evaluate the future demand and attempt to reach their charging goal - a sufficient state of charge (SOC) that would allow them to fulfil the next day's desired trip. At the same time they attempt to minimize the charging cost, which is directly proportional to the power demand in a real-time pricing mechanism. 

While such problems are relatively simple from a single RL agent's perspective - as eventually it will reach optimal performance - not the same can be said in an environment where agents compete against each other as well; once a lot of agents decide to charge at the same time, the price significantly increases, leading the agents to decide to charge at other points in time. This can lead to an endless loop, where distracted agents take the same charging point decision at each following timestep, and leads to suboptimal behaviour, as shown by \cite{Karfopoulos2013}.  This is a significant obstacle faced by the convergence of the MAS to a desired behaviour. 

A summary of P-MARL's decision process is presented in Alg. \ref{alg:p-marl}. The following sections present how P-MARL is implemented in the smart grid scenario, with focus on each of the three components. 
 
\begin{algorithm}
\begin{algorithmic}
\State $initVars = gatherEnvData()$
\State $prediction = initialPrediction(initVars)$
\If{$noChangeDetected()$ }
\State $finalPrediction = prediction$
\Else 
\State $matchChangeType()$
\State $input = updateInitVars()$
\State $finalPrediction = anomalyReprediction(input)$
\EndIf
\State$learnBestBehaviour(finalPrediction)$
\State{$exploitLearnedInfo()$}
\caption{P-MARL Algorithm}
\label{alg:p-marl}
\end{algorithmic}
\end{algorithm}

\subsection{1. The Prediction Model} 
Electric vehicles generally arrive home at 18:00 and depart he next day at 09:00. Ideally the agent would like to know the periods of lowest demand occurring during that time in order to be as price effective as possible. In a dynamic environment though such \emph{a  priori} knowledge is not avaialable. 

Techniques such as short-term load forecasting (STLF) deal with power demand estimates, which can give good hints on expected demand. The most appropriate such STLF sub-technique is the one focusing on day-ahead demand forecasting, on a 24 hour basis. The best methods for such forecasts rely not only on historical values of previous power demands but also on other data such as weather variables (temperature, humidity), day of the week and public holidays \cite{Gross1987}. While weekdays and weekends differ significantly in terms of demand patterns, even each weekday has it's own particularities. Special cases occur as well. These could be for example public holidays, so all such cases need to be taken into account. 

Our prediction model is implemented based on the work done in \cite{Marinescu2014a}. Here the forecasting method is a hybrid solution exploiting the best features of several forecasting techniques: artificial neural networks, neuro-fuzzy networks and auto-regression.
The hybrid solution uses as input previously recorded power demands, past day's temperature and humidity information, temperature and humidity forecasts for the day to be predicted, and day of the week information. The output is the next day's power demand estimate provided as a sequence of 24 data points, one for each hour of the day.

\subsection{2. The Pattern-Change Detection and Matching Component}

When dealing with uncertainty one can never have quality of service guarantees. Despite the fact that the previous model provides rather accurate predictions, there are particular times when forecasts fail to closely match actual demand. Such particular cases are caused by anomalous events such as electrical grid malfunctions,
unexpected climate phenomena or natural disasters. In order to make the system more robust in the face of such events an additional pattern changed detection component has been added. 

This component builds upon the work carried out in  \cite{Marinescu2014b}. Once the prediction model provides a forecasting estimate for the next day, the actual demand is compared with the estimate along the first few hours of the morning. If significant deflections from the actual demand occur, the pattern change detection mechanism triggers reprediction as the demand estimate is regarded as flawed. The re-prediction part is based on an artificial neural network which adjusts its historical load input part (24 neurons, one for each hour of the day) to a mash-up between the demand observed so far of the anomalous day (7:00-14:30) and the remaining demand from the closest match. The match is chosen based on similarly previously encountered patterns, which are found in a database of historical recordings - this function is carried out by a classifier involving a self-organizing map. The self-organising map first classifies the type of uncertainty detected and afterwards provides the closest previously such encountered match from the fitting class. 

\subsection{3. The Multi-Agent Reinforcement Learning System}

The previously proposed components are designed to be linked with a MARL, thus resulting in P-MARL. Each agent is developed individually by following a reinforcement learning scheme - in this case Q-Learning \cite{Watkins1992} combined with W-Learning. 

Initially this is a single-objective problem for each agent: to make sure that the desired charge is reached, reward the agent when charging. This results in a greedy behaviour, with the agent charging at every time-step until it's fully charged. Obviously this would lead to an undesirable behaviour, as the EV can easily charge during periods of high demand. The constraint of avoiding charging at periods of high demand is introduced under the form of another objective. This effectively transforms our problem into a multi-objective one. The second objective rewards the agent if it decides to charge at periods of low demand, and penalises it if it decides to charge at times of high demand . The second objective is highly relevant - it can be further modelled to employ information about the future state of the environment, which is provided by the prediction components. Essentially the prediction component provides a form of reward shaping for the agent through the extra-objective, but this is just an estimate, thus with no guarantees of certainty. If the estimate is good, P-MARLs results will be good; if the estimates are bad, P-MARL will perform sub-optimally. While still uncertain, we argue that a (good) estimate is still better than no estimate at all.

\section{Experimental Setup}

In this paper we apply P-MARL to an environment which is inherently dynamic, a real-world scenario occurring in the smart-grid. The state of the environment is characterised by its energy consumption, which involves randomness due to human users. The environment can be represented as a time-series defined by the half-hourly power demand. The time-series experiences a clear concept drift. 
By involving a set of controllable loads on top of a baseload demand (where the baseload is the aggregated demand of inflexible appliances such as lights, television sets or computers, which are exclusively controlled by humans) we reach a situation involving a group of agents with certain objectives to be accomplished, under some predefined constrains. In this case we have a neighbourhood of residential users which contains a set of EVs; the task of each EV is to achieve a desired state of charge (SOC) for the next day's trip. Additionally, this charging process might be constrained by periods of high demand, when electricity is expensive, and when charging is to be avoided. Such periods can change in real-time, for example when all EVs charge simultaneously during a period of relatively moderate energy usage, resulting in very high energy usage. The latter of course is not a desired state of the environment, and ideally should be avoided. 
\begin{figure}[!t]
\centering
\includegraphics[width=\linewidth]{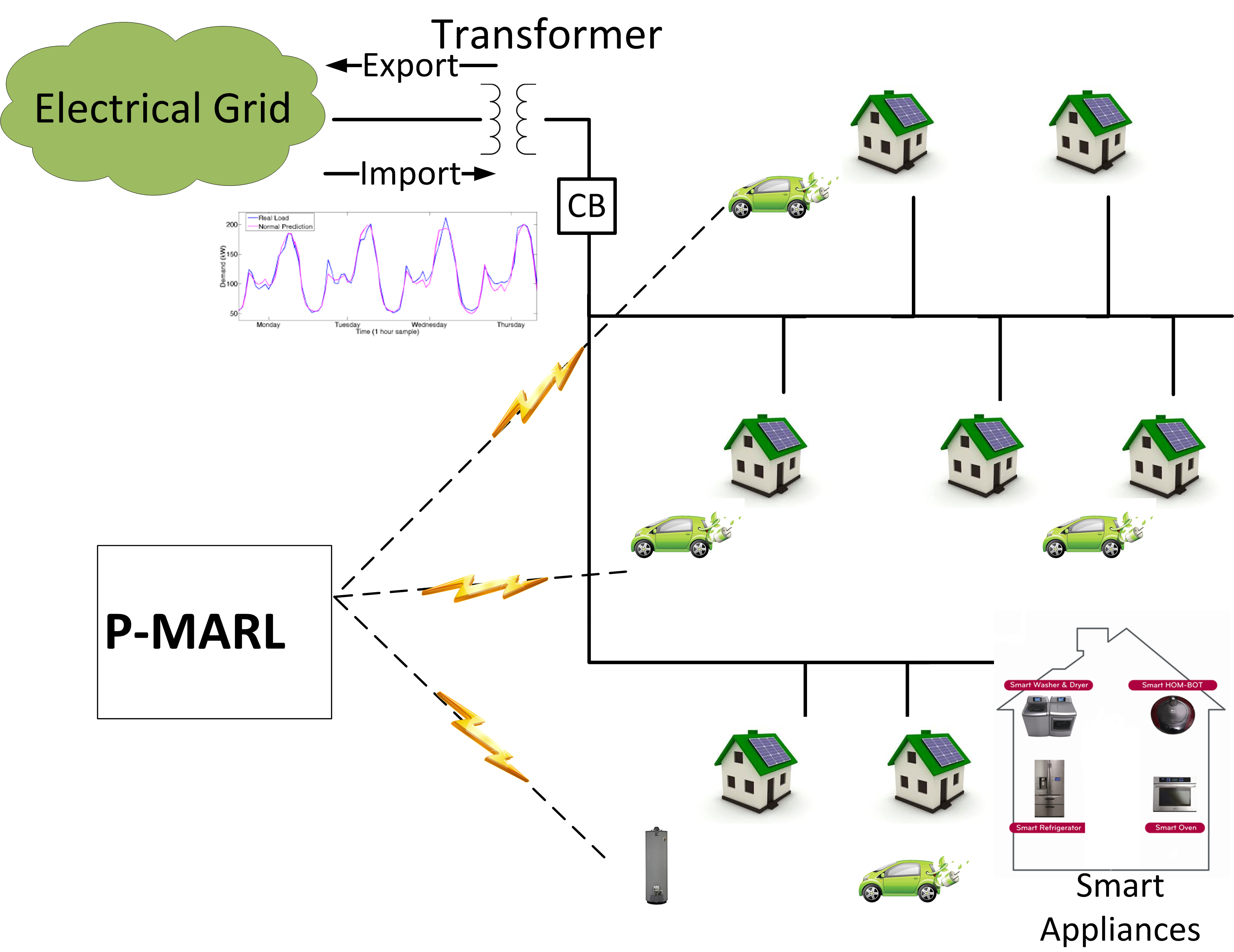} 
\caption[c]{Smart Grid Scenario}
\label{fig:grid_struct}
\end{figure}
The smart grid setup can be visualised in Fig. \ref{fig:grid_struct}. The charging algorithms are evaluated in a real world scenario. Power demands from a community of 230 households are employed, as recorded by an Irish smart meter trial \cite{CER2011}. We have assumed a penetration of EVs of 40\% \cite{Nemry2010}, resulting in a total of 90 EVs. Daily trip is considered to be 50\hspace{2pt}km \cite{EPA2008}, while the EV specifications are borrowed from \cite{Nissan2014}. The vehicles can choose charging slots anytime between 18:00-09:00. This smart grid scenario was implemented in GridLAB-D, a power distribution system simulator \cite{GridLABD2014}.

\begin{figure*}[!ht]
\centering
\subfigure[Perfect Prediction]{
\includegraphics[width=0.31\linewidth]{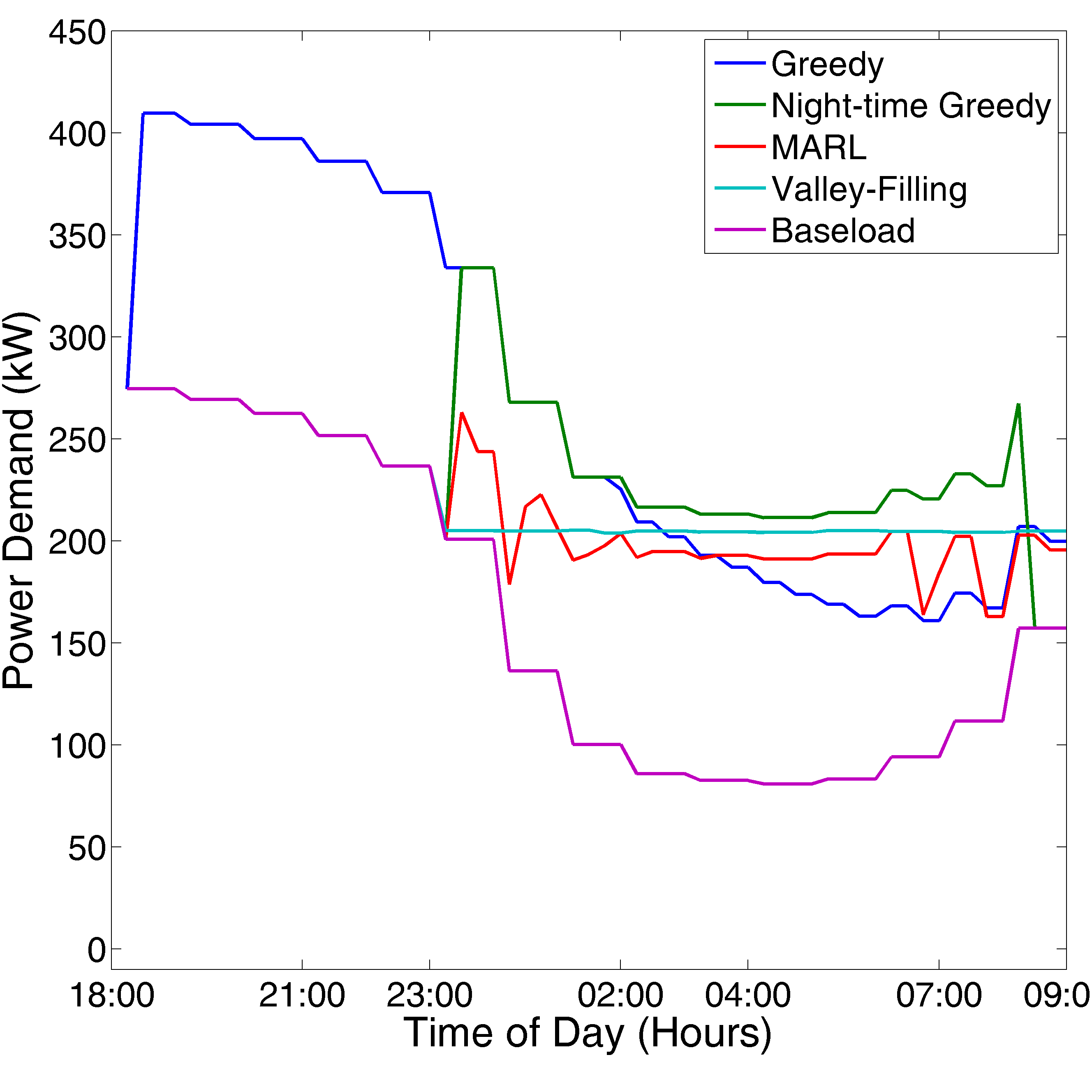} 
\label{fig:perfect_case}
}\quad
\subfigure[Anomaly Prediction]{
\includegraphics[width=0.31\linewidth]{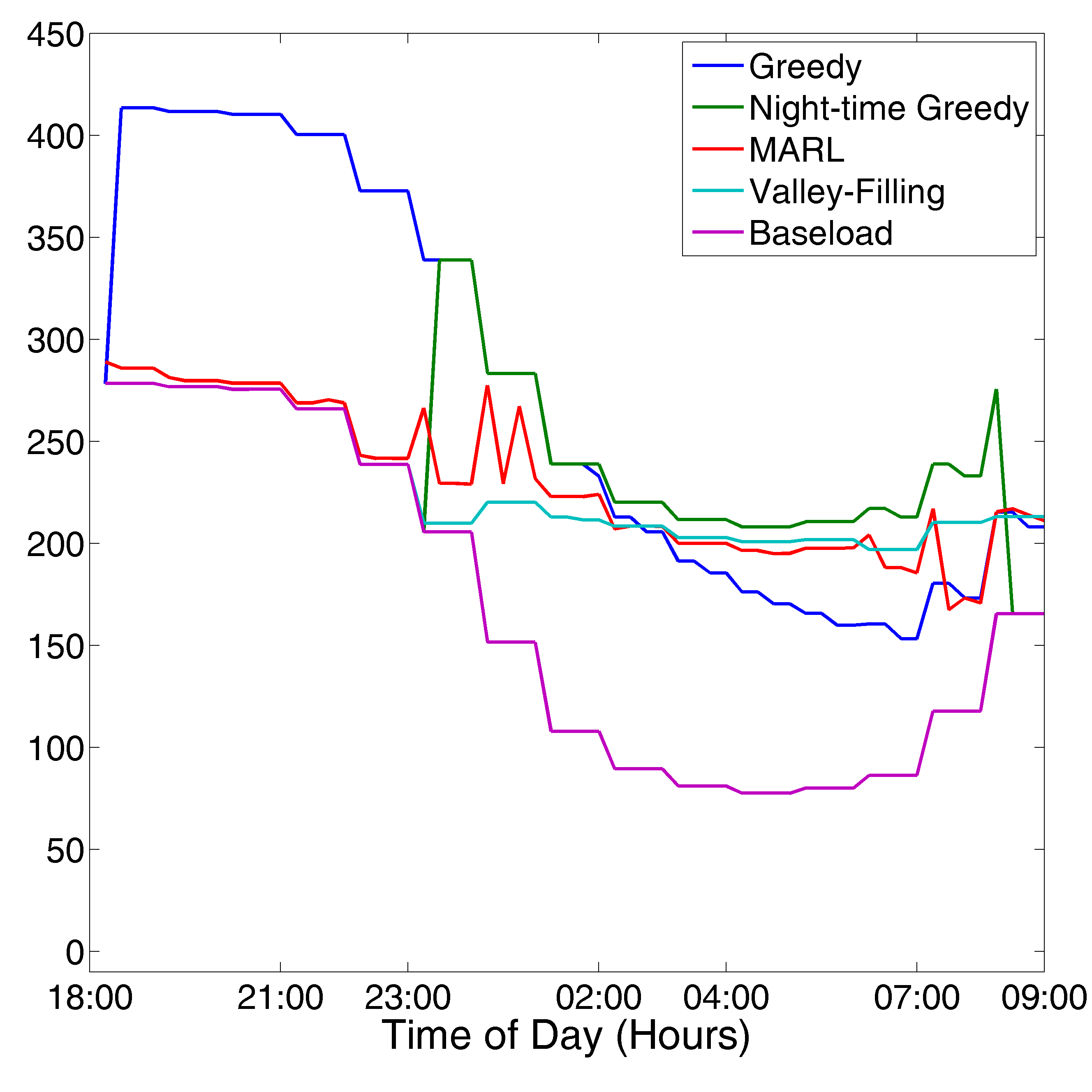}
\label{fig:apred_case}
}\quad
\subfigure[Simple Prediction]{
\includegraphics[width=0.31\linewidth]{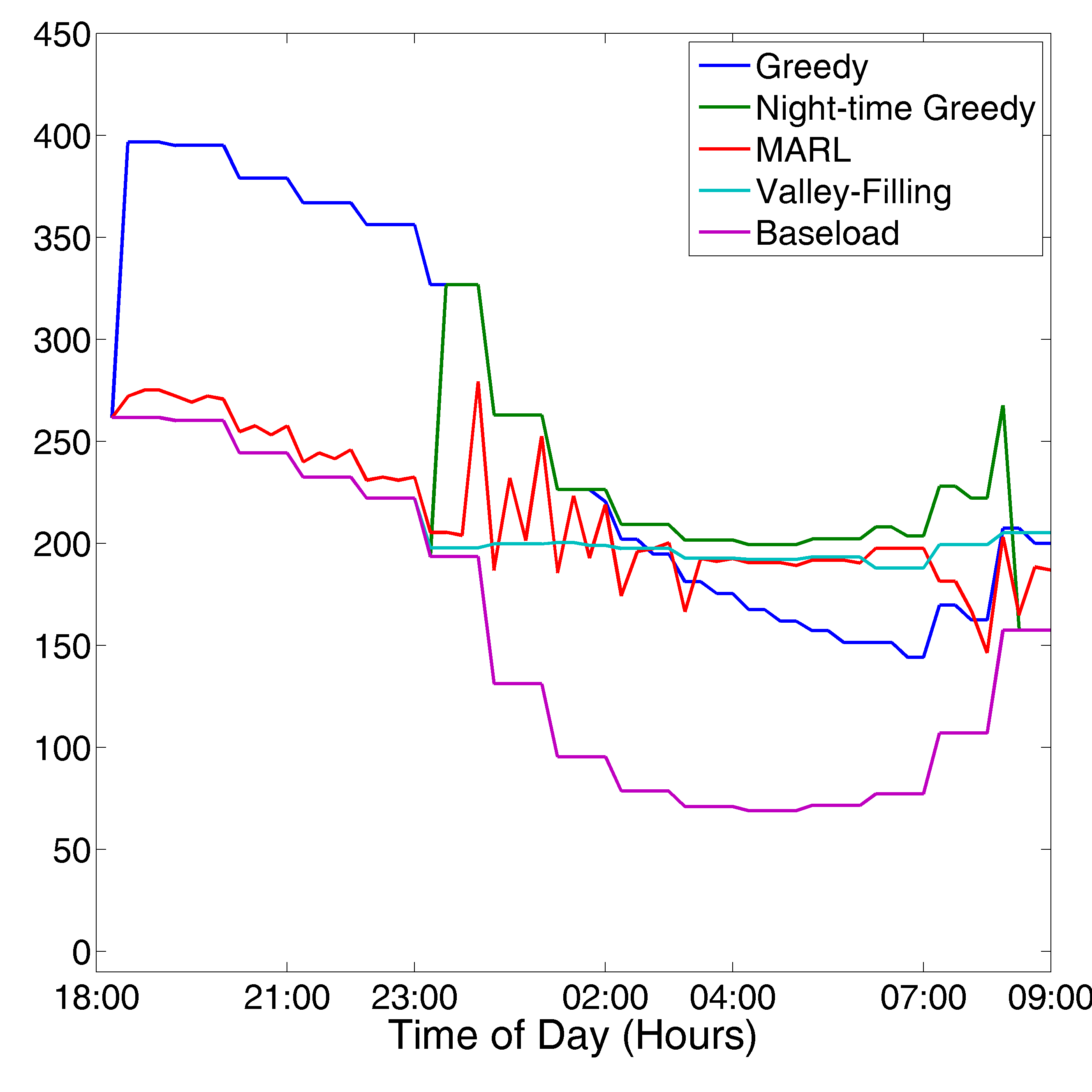} 
\label{fig:spred_case}
}
\caption{Algorihtms Behaviour}
\label{fig:algorithm_behaviour} 
\end{figure*}

\subsection{Benchmark - Optimal Centralised Solution}
In order to efficiently evaluate P-MARL we have to define the Pareto optimal performance of the MAS. As outlined in the previous section, a centralised solution is not suitable in such cases. Nevertheless, it's guaranteed to be optimal with respect to a defined set of constraints. Assuming a system of dynamic pricing, the solution should lead to EVs energy usage at the lowest demand times (given that energy cost is directly proportional with the system power load). The resulting constrained optimization function is presented in Eq. \ref{eq:func}:

\begin{equation}
\centering
\min F(x) = \min \sum\limits_{j=1}^m\left[\sum\limits_{i=1}^n\bigg(x_{ij}+C_{j}\bigg)\right]x_{ij}
\label{eq:func}
\end{equation} 

where $F(x)$ is the cost function, $n$ the total number of electric vehicles, $m$ the total hours available for charging (assuming the same availability schedules for EVs), $x_{ij}$ the charging decision (0-off/1-on) of vehicle $i$ at time $j$, and $C_{j}$ the initial cost of energy at time $j$ (based on baseload).

While solutions are achievable, the purpose of the benchmark is not to obtain individual solutions for each agent but to define the aggregated optimal charging solution of the overall vehicles. This is essentially a valley-filling problem, such as the one presented in \cite{Gan2011}.

 Ideally the MAS solution should converge to the same result as the centralised solution. In order to evaluate the efficiency of the MAS solution we used a formula based on the mean absolute percentile error (MAPE), as shown in Eq. \ref{eq:eff}.

\begin{equation}
M=\frac{1}{m}\sum\limits_{j=1}^m\bigg(1-\frac{|X_{j}-\hat{X_{j}}|}{TotalNoOfEVs}\bigg)
\label{eq:eff}
\end{equation}

where $X_{j}$ total number of EVs charging at time slot $j$, and  $\hat{X_{j}}$ optimal amount of EVs that should be charging at time slot $j$.

It is worth noting that one metric is not enough to show the overall performance of the system. The number of high deviations (above a certain threshold, say 25\%) from the desired optimal solution should also be used, as these can have quite an impact on the physical power networks, in particular at times of high energy usage.

\section{Evaluation}

The experiments are performed on the scenario mentioned before. Four different algorithms are evaluated:
\begin{itemize}
\item \emph{Greedy} Solution - which charges the EVs as soon as possible - unconstrained single-objective MARL
\item \emph{Night Tariff-Aware Greedy} solution - which charges the EVs as soon as possible starting from 23:00, by adjusting to a night-saver time tariff
\item \emph{Optimal} Valley-Filling (V-F) Solution -  obtained by the centralised algorithm
\item \emph{P-MARL} solution - given by our multi-objective algorithm presented in the System Architecture section \ref{sec:syst_arch}
\end{itemize}

At the beginning of the experiments the state of charge of each vehicle is randomly initialised with values in between [0.17-0.67]. The experiments are split over 3 different sub-cases: assuming the normal simple prediction of the day as an input, the more accurate re-prediction of the day (anomalous case), and finally assuming perfect forecast of the day. The later one is used just for comparison purposes, in order to see the effects of forecasting accuracy over algorithm performance. The algorithm continuously learns over a period of 100 days (same training day repeated 100 times). The experiments are run 10 times and averaged.

\subsubsection{Simple Prediction}
 
This sub-case is employed in order to show the difference between accurate daily forecasts and ones that are affected by a higher amount of forecasting errors, in particular in the situation of anomalous days. This can be visualised in Fig. \ref{fig:spred_case}.
  
\subsubsection{Anomaly Prediction Case}

This sub-case tests the algorithm on a relatively accurate forecast of the baseload demand. In non-anomalous days simple re-prediction achieves similar levels of accuracy, so it could be said that for normal days the other two sub-cases are sufficient in order to show effects of forecasting accuracy. We present the results obtained by the four algorithms in such instances in Fig. \ref{fig:apred_case}.   

\subsubsection{Perfect Prediction Case}

Here we assume that our estimation of the environment's future baseload demand is 100\% accurate. Obviously this is not the case in real life, but we use the preliminary results to see the effects of forecasting errors on algorithm performance.  The results of the four algorithms in this case are presented in Fig. \ref{fig:perfect_case}.

\subsection{Results and Analysis} 

\begin{figure}[!t]
\centering
\includegraphics[width=0.9\linewidth]{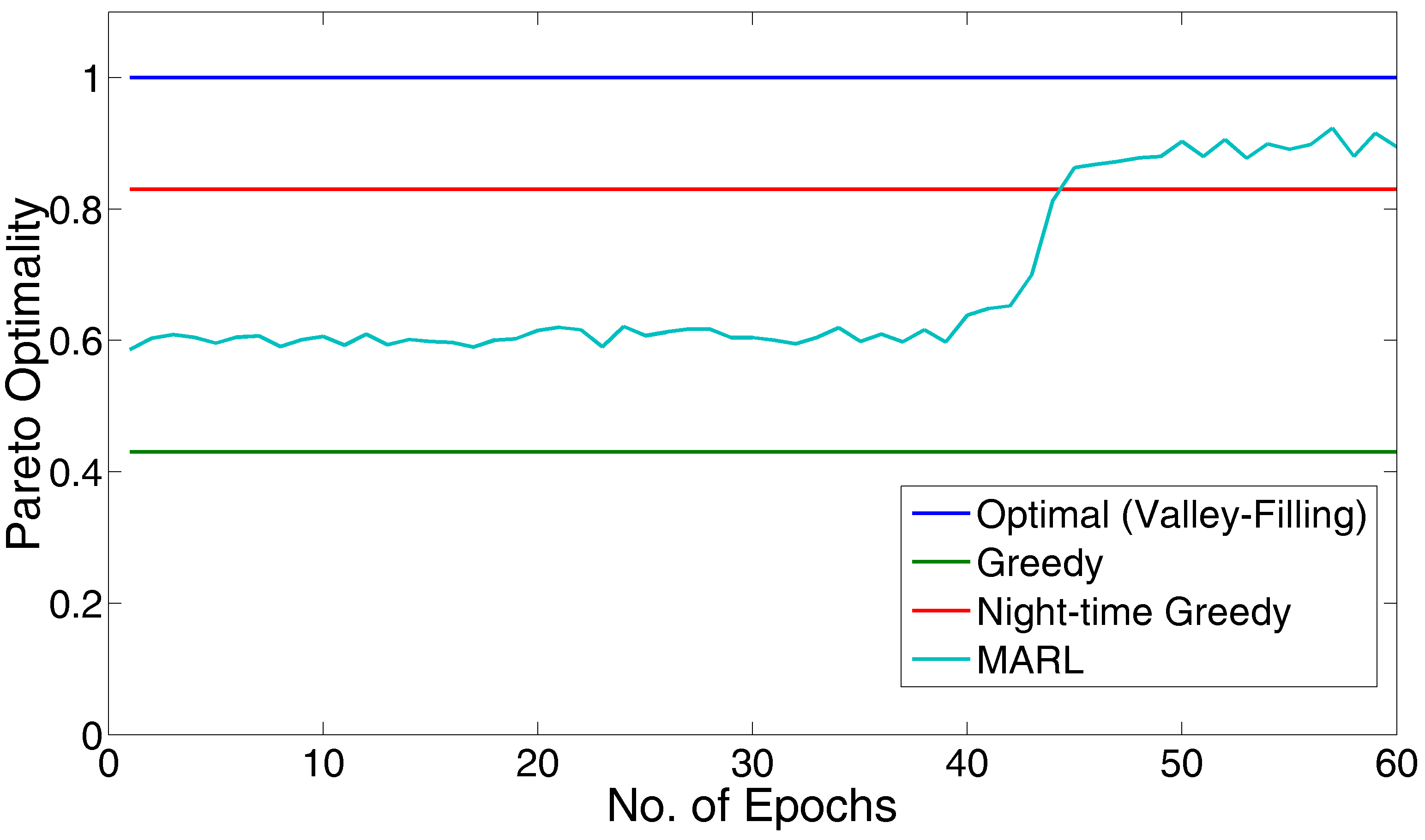} 
\caption[c]{Efficiency Evaluation}
\label{fig:perfect_eff}
\end{figure}

From the graphs we can see that the simple Greedy solution performs worst when compared with the other three methods. This is quite obvious once we look at the efficiency graph in Fig. \ref{fig:perfect_eff}. As soon as the vehicles arrive home they start charging, which occurs exactly during the peak time consumptions (which is the expected behaviour when people come home and start using their home appliances such as ovens, kettles, washing machines, lights, etc.). As a result EVs cause almost an increase of 50\% in demand, which has to be accounted for by turning on additional power plants (expensive). The improved greedy solution leads to the vehicles deciding to charge at night time, starting from 23:00. While this doubles the performance of the Greedy solution (as seen in Fig. \ref{fig:perfect_eff}), if we take a closer look on the actual impact over the power demand curve from Fig. \ref{fig:perfect_case}, we can notice that it also has an undesired effect. The demand suddenly peaks at 23:00 creating a spike significantly higher than the baseload demand at peak time. Again this creates additional problems on the generator side of power networks and leads to unnecessary additional costs. A somewhat similar pattern can also be noticed when analysing the behaviour of the MARL algorithm. As soon as demand decreases several EVs choose to start charging. As the aggregated amount of energy used increases, the EV agents realise immediately that their choice is leading to an undesired state and start backing off. The follow-up is a usage pattern that is much closer to the desired optimal valley-filling behaviour achieved by the centralised algorithm. There are still some random variations which happen due to the same simultaneous decision of several agents. Fig. \ref{fig:perfect_eff} points out the evolution of the MARL as it moves from exploration to exploitation stages. It explores various possibilities for 40 learning episodes (days), and afterwards it starts exploiting. This is clearly noticeable in the graph. We have a sudden jump in efficiency, and then algorithm requires only 10 more learning episodes to converge to it's near-optimal solution (a 92\% Pareto optimal solution). 
\begin{table}[b]
\small
\caption{Comparison of Performance}
\begin{tabular}{|l|c|c|c|c|}
\hline
\bf{Method} & \bf{Greedy} & \bf{N. Greedy} &\bf{MARL} & \bf{V-F} \\ 
\hline
\bf{Perfect} & 46.1\% & 83.1\% & 92.4\% & 100\%\\ 
\hline
\bf{Repredicted} & 44.1\% & 83.5\% & 92.2\% & 97.6\%\\ 
\hline
\bf{Simple} & 44.6\% & 83.8\% & 89.6\% & 97.9\%\\ 
\hline
\end{tabular}
\label{tab:perf_anal}
\end{table}
The greedy solutions are obviously inferior in terms of Pareto optimality and effects on the electrical grid, but they also choose to charge the vehicles more than the other algorithms. While this can be useful in unexpected situations, on average the cars will have a higher SOC at the beginning of the day than required for their daily trips. The valley-filling algorithm will always charge precisely the amount required for the daily trip, while the MARL algorithm shows a somewhat more sensible form of reasoning: if the car is generally charged enough (80\%), the EV agents tend to be satisfied and some stop charging even though they would have more available charge slots. This behaviour does not show any bad effects on the EV performance, as after ten experiments each worth 60 learning episodes, the vehicles have never even once run out of charge. Moreover, they all arrive home with a SOC in between 40\%-60\% - which obviously leaves them space for those extra unexpected trips. Another side effect is the fact that battery life is therefore expected to increase if the battery is not fully charged each time \cite{Hoffart2008}.

The performance of the algorithms in terms of Pareto optimality is summarised in Table \ref{tab:perf_anal}. As expected the best performance is achieved in the perfect prediction case, assuming perfect anticipation of the future power demand. Even in the case of reprediction the performance of the MARL is only minimally diminished (0.2\%) as it is able to adjust in real-time, unlike the other algorithms which are schedule based. This can be seen in Fig. \ref{fig:apred_case}. The Greedy and Night-Greedy algorithms have very similar performance in all three cases. Together with the decrease in forecasting accuracy comes a decrease in the performance of P-MARL as well. The increase in forecasting errors from 4.66\%\hspace{2pt}MAPE in the anomalous reprediction method  to the 7.66\%\hspace{2pt}MAPE of the simple prediction comes with a price in terms of MARL performance, which is brought down from 92.4\% to 89.6\% Pareto optimality. P-MARL  still outperforms the greedy solutions, while the centralised solution's performance doesn't seem to be affected at all in this case. This is due to the fact that, despite significant differences in demand (the initial prediction provides an estimate whose peak is underestimated by 5\%) the actual curve's shape maintains the same pattern in this particular case, thus allowing the valley-filling algorithm to perform similarly. P-MARL's second objective is based on statistics computed from the estimated demand (average demand and standard deviation), thus lower estimated peaks lead to EV power usage in times of actual high demand of the day, as can be noticed in Fig. \ref{fig:spred_case} where some EV charge occurs even during peak times. The EVs expect periods of higher demand to occur later, so they try to take advantage of the periods of supposedly lower demand. This leads to an undesired effect. A small percentage of the EVs actually end up charging in the period with the highest demand, thus the least cost-effective one.

\section{Conclusions and Future Work}

This paper presents P-MARL, a solution which improves MARL performance in uncertain environments. The proposed solution reaches near-optimal results, within 92\% Pareto optimality. As noticed in the Evaluation section, the performance of the P-MARL  is closely related to the ability of being able to accurately forecast future states of the environment. The loss in accuracy results in diminished performance, therefore the need of very good environment prediction mechanisms is justified. Uncertainty can be dealt with by pattern-change detection elements which are able to trigger the re-evaluation of the environment's future behaviour. The effects of such a mechanism on forecasting accuracy and algorithm's performance has been noticed in the experimental section of the paper. Even though assuming perfect estimation of the environment's future behaviour, the dynamics implied by multi-agent systems lead to stochastic behaviour resulting sometimes in undesired effects. In this scenario we have proposed P-MARL as an online non-collaborative MARL solution, with no possibilities of communication. If we disregard the privacy concerns and communication overhead of collaborative MARLs, we believe that agents' cooperation in collaborative environments should lead to improved solutions, without significant deflections from the optimal performance. This will be the scope of future investigations, where we intend to apply collaborative algorithms such as DWL \cite{Dusparic2010}.

\newpage

\bibliographystyle{aaai}
\bibliography{aaai}

\end{document}